Authors:
  Takano, Y., Chikaraishi, Y. and Ohkouchi, N. (2010)

Title:
  Enantiomer-specific isotope analysis (ESIA) of D- and L-alanine:
  nitrogen isotopic hetero- and homogeneity by microbial process and
  chemical process.

  **Earth, Life, and Isotopes** (edited by N. Ohkouchi, I. Tayasu, and K. Koba).

  Kyoto University Press., pp. 387-402.

  --
  Please see also,
    http://www.kyoto-up.or.jp/book.php?lang=en

**Enantiomer-specific isotope analysis of D- and L-alanine: Nitrogen isotopic hetero- and homogeneity in microbial and chemical processes**

Yoshinori Takano[*], Yoshito Chikaraishi, and Naohiko Ohkouchi

Institute of Biogeosciences,

Japan Agency for Marine-Earth Science and Technology (JAMSTEC),

2-15 Natsushima, Yokosuka 237-0061, JAPAN

*Corresponding author

Yoshinori Takano

Fax: +81-46-867-9775

E-mail: takano@jamstec.go.jp

**Abstract**

Nitrogen isotopic hetero- and homogeneity of D-α-alanine and L-α-alanine were investigated in microbial processes in the domain Bacteria and in chemical processes in symmetric organic synthesis. D-alanine is an enantiomer that is physiologically essential for microbial growth and metabolic maintenance. The nitrogen isotopic difference $\Delta^{15}N_{D-L}$ (defined as $\delta^{15}N_{D-Ala} - \delta^{15}N_{L-Ala}$) in peptidoglycan amino acids in bacteria such as the representative gram-positive phyla *Firmicutes* and *Actinobacteria* (*Enterococcus faecalis*, *Staphylococcus aureus*, *Staphylococcus staphylolyticus*, *Lactobacillus acidophilus*, *Bacillus subtilis*, *Micrococcus luteus*, and *Streptomyces* sp.) tended to be $^{15}N$-depleted in D-alanine ($\Delta^{15}N_{D-L} < -2.0‰$ ). These results suggest that the composition of isotopically heterogeneous components in these bacteria is primarily controlled by enzymatic pathways prior to formation of the bacterial cell wall. In contrast, the $\Delta^{15}N_{D-L}$ of racemic alanine in the chemical pathway during the nucleophilic substitution reaction (SN1 type) between 2-bromopropionic acid and ammonia identified fully homogeneous components for each enantiomer. The novel enantiomer-specific isotopic analysis (ESIA) method is useful in determining the origins of chirality in biogenic and abiogenic processes and is applicable to enantiomer studies.

**Introduction**

The one-handedness of terrestrial L-amino acids in the proteins and D-sugars of DNA and RNA are essential to the formation, structure, and function of biopolymers for life on Earth. From the middle of the 19[th] century onward, the putative origins of biomolecular homochirality have been abundantly reported in the literature from both biotic (biological) and abiotic (chemical and physical) viewpoints (e.g., Bonner 1991, 1995; Palyi et al., 1999). Since the initial analysis of pristine meteoritic amino acids in the 1960s (e.g., Hayes, 1967; Kvenvolden 1970; Lawless et al., 1971; Oro et al.,1971; Cronin et al., 1979; Shimoyama et al., 1979), abiotic exogenous amino acids have been recognized as differing from biotic endogenous terrestrial amino acids (e.g., Cronin and Pizzarello 1997; Kvenvolden 2000; Sephton 2002). Among the D- and L-amino acids found in nature, α-alanine (Ala) is the simplest enantiomer, the L-isomer of which is one of the 20 proteinogenic amino acids. Although meteoritic D- and L-alanine have been separated by gas chromatography following chemical derivatization, important aspects of their behavior relating to enantiomeric excess and compound-specific nitrogen isotopic composition remain unconfirmed (see Engel and Macko 1997, 2001; Pizzarello and Cronin 1998, 2000). Previous studies have reported that the stationary phase containing nitrogen compounds should be used with caution for the precise determination of $^{15}N/^{14}N$ ratios (Metges and Petzke, 1999). For instance, Macko et al. (1997) and Chikaraishi et al. (2010) noted a potential inconsistency in the $\delta^{15}N$ of amino acids, possibly resulting from a minor contribution of nitrogen from the stationary phase.

55    Although L-amino acids are predominant in the proteins of living organisms, D-amino acids including D-alanine, D-glutamate, and others also occur in members of the domain Bacteria incorporated into peptidoglycan (e.g., Schleifer and Kandler, 1972), which is an essential and specific component of the bacterial cell wall and found on the outside of the membrane of most bacterial species (van Heijenoort, 2001a, b; Vollmer et al., 2008). The cell

60    walls of gram-positive bacteria consist of a thick uniform peptidoglycan layer that includes D-amino acids, which forms up to 90% of the cell wall. Alanine racemase catalyzes interconversion between L-alanine and D-alanine using pyridoxal 5'-phosphate (PLP) as a coenzyme, and plays a central role in D-alanine metabolism. These enzymes support the growth of bacteria by providing D-alanine, which is a crucial molecule in peptidoglycan

65    assembly and cross-linking (e.g., Schleifer and Kandler, 1972; Wasserman et al., 1984; Walsh 1989; Hols et al., 1997; Nagata et al., 1998; Yoshimura and Esaki, 2003). Bacterial culture surveys of the deep-sea subsurface environment have found a predominance of these gram-positive bacteria in benthic microbial communities (e.g., D'Hondt et al., 2004; Teske, 2006). On the basis of these culture surveys, D-amino acids in the peptidoglycan layer are clearly preserved even in high-pressure sub-surface environments. Gram-positive bacteria and hypotheses for their emergence have also been addressed in molecular phylogenetic studies in the context of the search for a universal root of the phylogenetic tree (Lake et al., 2007; Skophammer et al., 2007).

To date, most investigations of amino acid enantiomers have been performed by examining enantiomeric excess (%D – %L) or chiral ratios (D/L). Here, we report on a newly established methodology employing enantiomer-specific isotopic analysis (ESIA) using an online gas chromatograph/combustion/isotope ratio mass spectrometer (GC/C/IRMS).

To date, two chromatographic methods for chiral separation have been commonly used (Schurig, 2001). The first, enantiomer separation using the chiral stationary phase, is a typical procedure. This method has been developed for stable isotope studies of D- and L-amino acids using gas chromatography (GC) techniques with chiral polysiloxane (Chirasil-type) stationary phase columns with a valine diamide selector (Frank et al., 1977), later known as Chirasil-Val. The second is diastereomer separation, in which chemical derivatization with optically active reagents is performed without a chiral stationary phase. In the present study, we examined nitrogen isotopic imbalances using ESIA (Figure 1) for $\delta^{15}N_{D\text{-}ala}$ and $\delta^{15}N_{L\text{-}ala}$ to determine their biogenic and abiogenic (pristine organic chemical

processes in the absence of biological processes) origins by using an improved diastereomer

separation method. Employment of ESIA in determination of $^{15}N/^{14}N$ ratios has the potential

to assist in delineation of D- and L- amino acid behavior in biological and chemical processes

90    to a greater extent than compound-specific isotopic analysis (CSIA) studies.

## Materials and Methods

### *Microbial samples*

Figure 2 shows a cross section of the structure of bacterial peptidoglycan containing

95    D-amino acid peptide linkages. Purified peptidoglycans were used from the domain Bacteria

(phyla *Firmicutes* and *Actinobacteria*; *Enterococcus faecalis*, *Staphylococcus aureus*,

*Staphylococcus staphylolyticus*, *Lactobacillus acidophilus*, *Bacillus subtilis*, *Micrococcus

luteus*, and *Streptomyces* sp.), as well as (pseudo)-peptidoglycans from the domain Archaea

(*Methanobacterium* sp.), cell walls from domain Eukarya (*Saccharomyces cerevisiae*), and

100    occasionally whole cells from natural *Bacillus subtilis* var. *natto* for bulk cell analysis following the elimination of extracellular polysaccharides (Takano et al., 2009).

*Chemical samples: abiotic synthesis of D- and L-alanine*

Racemic D- and L-alanine were synthesized through a nucleophilic substitution 1 (SN1) reaction via an intermediate carbocation formed from α-bromopropionic acid and aqueous ammonia (Figure 3). The synthetic procedure used was a modification of published methods (Tobie and Ayres, 1937; Pine, 1987). In brief, 10 g of cold (1–4 ℃) α-bromopropionic acid was slowly poured into 200 g of concentrated aqueous ammonia (25% $NH_{3aq}$) with stirring. The mixture was then held at room temperature (below 40 ℃) in a draft chamber. After at least 4 days reaction time, the mixture was evaporated under nitrogen flow to 10 mL. Thirty mL methanol were added and the mixture was cooled overnight at 1-4 ℃ for recrystallization. The obtained crystals were filtered using a GF/F filter, carefully rinsed with methanol and diethyl ether, and finally dried at ambient temperature.

To determine the identity and purity of the product, 25 mg of the product was dissolved in 0.5 mL of $D_2O$ (> 99%) in a 5-mm sample tube and subsequently analyzed by [1]H-NMR and [13]C-NMR (400 MHz; Bruker BioSpin, Billerica, MA). The purity, D/L ratio, and presence of byproducts were determined by GC/MS (6890N/5973MSD; Agilent Technologies, Santa Clara, CA) through screening of the reaction products following the scheme shown in Figure 3. The GC capillary column was an HP-5ms (30 m x 0.32 mm i.d.,

120    0.52 μm film thickness; Agilent Technologies). The GC oven temperature was programmed

as follows: initial temperature of 40°C for 4 min, ramped up by 10°C min$^{-1}$ to 90°C, and then

by 5°C min$^{-1}$ to 220°C, where it was maintained for 10 min. The MS was scanned over an

*m/z* of 50–550 with the electron-impact mode set at 70 eV. The GC/FID (Agilent 6890) was also equipped with a HP-5ms capillary column and was controlled by the same oven program,

125    as was the GC/MS.

***Enantiomer-specific nitrogen isotope analysis of D- and L-alanine***

We have previously developed an analytical method to conduct ESIA of individual

amino acid enantiomers (Takano et al., 2009). Novel derivatization of amino acid

130    diastereomers by optically active (*R*)-(−)-2-butanol or (*S*)-(+)-2-butanol with pivaloyl chloride produces N-pivaloyl-(*R,S*)-2-butyl esters (NP/2Bu) of the amino acid diastereomers,

which affords two advantages for ESIA. First, chromatographic chiral separation can be

achieved without the use of chiral stationary-phase columns and can be conducted at high

temperatures: for example, the Ultra-2 capillary column (Agilent Technologies) has a

135    temperature range of up to 325°C. Second, the elution order of the compounds on the chromatogram can be reversed by a designated esterification reaction (Takano et al., 2009).

In the present study, the nitrogen isotopic composition of the individual amino acids was determined using a GC/C/IRMS with a Delta Plus XP (Thermo Finnigan, Austin, TX) combined with a 6890N GC (Agilent Technologies) and an Ultra-2 capillary column (5% phenyl 95% methyl polysiloxane, 25 m x 0.32 mm i.d., 0.52 μm film thickness; Agilent Technologies; see details in Chikaraishi et al., 2010) in combustion and reduction furnaces (Chikaraishi et al., 2007; Takano et al., 2009). Combustion was performed in a micro-volume ceramic tube with CuO, NiO, and Pt wires at 1000°C. Reduction was performed in a micro-volume ceramic tube with a Cu wire at 550°C. The GC oven temperature was programmed as follows: an initial temperature of 40°C for 4 min, ramped up by 15°C min$^{-1}$ to 130°C, next ramped up by 1°C min$^{-1}$ to 160°C, and then ramped up by 30°C min$^{-1}$ to 260°C, where it was maintained for 10 min. The carrier gas (He) flow rate was 1.3 mL min$^{-1}$. The $CO_2$ generated in the combustion furnace was eliminated using a liquid nitrogen trap. The nitrogen isotopic composition is expressed as the per mil (‰) deviation from the standard (vs. air), as defined by the following equation: $\delta^{15}N$ (‰) = [($^{15}N/^{14}N$)$_{sample}$ / ($^{15}N/^{14}N$)$_{standard}$ − 1] x 1000.

**Results and Discussion**

***Biological processes related to D- and L- alanine: bacterial peptidoglycans and whole cells***

155       Figure 4 shows representative chromatograms of peptidoglycan samples derived from *Micrococcus luteus* (phylum *Actinobacteria*) and *Lactobacillus acidophilus* (phylum *Firmicutes*). The relative percent abundances of D-alanine [vs. total D,L-alanine; %D /(%D + %L) x 100] for *Micrococcus luteus* and *Lactobacillus acidophilus* were 23.7% ± 1.5% (n = 3) and 25.6% ± 0.1% (n = 3), respectively. D-amino acid metabolism, one of the critical

160      processes required for bacterial growth, is summarized in Figure 5. Although bound glutamine would be altered to glutamic acid following hydrolysis, thus far we have been unable to distinguish between glutamine and glutamic acid among the mixed hydrolysis products. Considering the impurities of derivative reagents (Takano et al., 2009), the amino acid diastereomers should be less than 2%. Furthermore, racemization of the amino acid

165      standards during the 22-h hydrolysis treatment ranged 0.5%-1.3% for D-alanine generated from L-alanine (Amelung and Zhang, 2001).

      D- and L-alanine separated by baseline resolution were analyzed for enantiomer-specific nitrogen isotopic composition using GC/C/IRMS (Takano et al., 2009). $\delta^{15}N_{D\text{-Ala}}$ for peptidoglycan tended to be depleted in $^{15}N$ relative to $\delta^{15}N_{L\text{-Ala}}$ in the bacterial

170      peptidoglycans and whole cells of *Bacillus subtilis* var. *natto* (Figure 6). For example,

D-alanine of *Staphylococcus staphylolyticus* and *Bacillus subtilis* was depleted in $^{15}$N relative

to L-alanine in both organisms by 2.0‰. D-alanine was not detected when

(pseudo)-peptidoglycans from *Methanobacterium* sp. (domain Archaea) or cell walls from

*Saccharomyces cerevisiae* (domain Eukarya) were analyzed using this GC/C/IRMS system.

175   D-alanine was correlated with $\Delta^{15}N_{D\text{-}L}$ (defined as $\delta^{15}N_{D\text{-}Ala} - \delta^{15}N_{L\text{-}Ala}$), except for *Bacillus*

*subtilis*, suggesting some form of enzymatic control of the incorporation process during the formation of glycan chains. The structure of in vitro polymerized peptidoglycans may differ from those formed in native in vitro processes (Kraus et al., 1985; van Heijenoort 2001a).

Alanine racemase (Enzyme Commission, EC 5.1.1.1), an isomerase that interconverts

180   L-alanine to D-alanine, has been identified by gene analysis in cultured bacteria (Table 1):

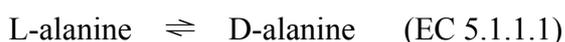
L-alanine   ⇌   D-alanine     (EC 5.1.1.1)

Furthermore, alanine racemase has been identified as a key enzyme in the D-alanine biosynthetic pathway (e.g., Hols et al., 1997), and participates in a crucial enzymatic reaction prior to the D-alanine-D-alanine ligase (EC 6.3.2.4) pathway in peptidoglycan metabolism.

185   Alanine racemase has been reported to be an essential limiting factor for the growth of some bacteria, including *Lactobacillus plantarum* (Wassermann et al., 1984). An additional pathway utilizes D-alanine transaminase (EC 2.6.1.21), which forms D-alanine and α-keto-glutarate from D-glutamate and pyruvate, and D-amino acid aminotransferase (EC

2.6.1.41), which forms D-alanine and 2-oxo acid from D-amino acid and pyruvate, as

190    follows:

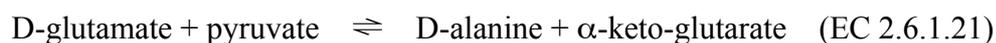

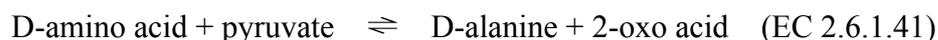

It is interesting to note that although most known bacteria possess alanine racemase,

some of these organisms possess no D-alanine transaminase even in culture inventories of

195    *Staphylococcus aureus*; in addition, *Enterococcus faecalis*, *Staphylococcus aureus*,

*Lactobacillus acidophilus*, and *Bacillus subtilis* were negative for D-amino acid aminotransferase in the present study (Table 1). These results suggest that enzymatic reaction pathways, primarily alanine racemase reactions, induce differences in the nitrogen isotopic composition of amino acid enantiomers.

200

***Chemical processes related to D- and L- alanine: abiotically synthesized products of the***

***SN1 reaction***

Pristine organic chemical synthesis in the absence of biological processes, or abiotic

synthesis, can produce symmetric organic compounds such as racemic D- and L-alanine

205 (Figure 3). The general form for this reaction may be given as:

$$R\text{-}LG + Nuc: \rightarrow R\text{-}Nuc + LG:$$

where the leaving group (LG) represents functional groups such as –Br, and $NH_3$ represents the nucleophile (Nuc).

The D/L ratio of synthesized alanine in the present study was estimated to be $1.000 \pm$

210 $0.002$ (n = 5), as determined by chiral separation using GC/FID. The purity was estimated to be > 99% based on a full-scan spectrum of $^{13}$C-NMR and $^{1}$H-NMR (Figure 3). The product yield as pure racemic alanine was > 77.6% based on the initial substrate and its products.

$\delta^{15}N_{D\text{-}ala}$ and $\delta^{15}N_{L\text{-}ala}$ were $-0.1 \pm 0.4$‰ and $-0.1 \pm 0.3$‰, respectively, as determined by

GC/C/IRMS. Consequently, the $\Delta^{15}N_{D\text{-}L}$ of synthesized alanine indicated complete

215 homogeneity ($0.0 \pm 0.4$‰) in the present study. We conclude that pristine organic synthesis in the absence of asymmetric catalysis and biological processes is capable of producing homogeneous isotopic compositions resulting from 50:50 symmetric substitution of nucleophiles in the SN1 reaction pathway.

220 ***Significance and applications of ESIA to studies of D- and L- alanine and other amino***

***acids***

L-Enantiomer excesses (0-15.2%) have been observed in extracted amino acids from the Murchison and Murray carbonaceous chondrites by GC/MS analyses of their N-trifluoroacetyl or N-pentafluoropropyl isopropyl esters, particularly for six

225    $\alpha$-methyl-$\alpha$-amino alkanoic acids (Cronin and Pizzarello, 1997; Pizzarello and Cronin, 2000; Pizzarello et al., 2003). Although enantiomeric excess and compound-specific nitrogen isotopic compositions of D,L-alanine in the Murchison meteorite remain unconfirmed (see Engel and Macko 1997, 2001; Pizzarello and Cronin, 1998, 2000), reported values for $\Delta^{13}C_{D\text{-}L}$ (defined as $\delta^{13}C_{D\text{-amino acid}} - \delta^{13}C_{L\text{-amino acid}}$) and $\Delta D_{D\text{-}L}$ (defined as $\delta D_{D\text{-amino}}$

230    $_{acid} - \delta D_{L\text{-amino acid}}$) are compiled in Figure 6.

Although careful evaluation is required of isotopic fractionation during the derivatization procedure in esterification and acylation (e.g., Silfer et al., 1991; Chikaraishi and Ohkouchi, 2010), the $\Delta^{13}C_{D\text{-}L}$ and $\Delta D_{D\text{-}L}$ for the Murchison meteorite indicate L-amino acid depletion for both $^{13}C$ and deuterium (e.g., Pizzarello et al., 2004; Pizzarello and Huang,

235    2005). The Murray meteorite also had depleted L-alanine in deuterium, with a $\Delta D_{D\text{-}L}$ for alanine of $104 \pm 61$‰ (n = 3) (Pizzarello and Huang, 2005). Thus, an undefined asymmetric isotopic fractionation process presumably yielded isotopic heterogeneity in terms of $\Delta^{13}C_{D\text{-}L}$ and $\Delta D_{D\text{-}L}$, implying that unknown physical and/or chemical processes occurred in the abiogenic stage prior to the meteorite reaching Earth.

240

**Conclusions and Perspectives**

(1) We report a cross section of nitrogen isotopic hetero- and homogeneity resulting

from microbial and chemical processes. For biological processes, we focused on D-alanine,

an essential enantiomer for bacterial growth. The $\Delta^{15}N_{D-L}$ of the alanine enantiomer tended

245    toward $^{15}$N-depletion in D-alanine, indicating heterogeneous components primarily controlled by enzymatic pathways. Investigation of other domain Archaea and Eukarya members (e.g.,

Corrigan, 1969; Yohda et al., 1996; Imai et al., 1997; Nagata et al., 1998) could be conducted

using the current ESIA method. For pristine chemical processes we verified an organic

chemical synthesis using nucleophilic substitution (SN1 type) and showed that completely

250    homogeneous components of each enantiomer had formed.

2) The ESIA method is applicable to various areas of life science research, including

food chemistry, pharmaceutical chemistry, and biochemical metabolomics (Man and Bada,

1987; Friedman, 1999; Leffingwell, 2003; Goodfriend et al., 2003; Kawai and Hayakawa,

2005). The bacteria examined in the present study include the *Actinobacteria* phylum and

255    *Firmicutes* phylum, which are ubiquitous in terrestrial soils and oceanic sediments of the sub-seafloor. Subseafloor biosphere culturing studies, one of the main focuses of the

Integrated Ocean Drilling Program (IODP), succeeded in isolating diverse members of

*Proteobacteria*, gram-positive bacteria (*Firmicutes* and *Actinobacteria*), and members of the

*Bacteroides* phylum (e.g., D'Hondt et al., 2004; Teske 2006). The most frequently obtained

260    cultured strains from an open ocean site and the Peru margin were members of the

spore-forming genus *Bacillus* within *Firmicutes* from an open ocean site and Peru margin

(Figure 7). Toffin et al. (2004) also reported similar phylogenetic groups, including enriched

members of *Firmicutes*, gamma- and delta-*Proteobacteria*, and *Spirochaeta*, during a

cultivation survey in Ocean Drilling Program Leg 190 sediments in the Nankai Trough. The

265    wide distribution of gram-positive, spore-forming bacteria may indicate resistance to nutrient depletion and physicochemical fluctuations, including changes in redox conditions and long time-scale sedimentary compaction.

**Acknowledgments**

270         The authors would like to thank Dr. K. Koba (Tokyo Univ. Agri. Tech) and an anonymous reviewer for their critical and constructive review comments, which helped improve the manuscript. This research was partly supported by a Japan Society for the Promotion of Science (JSPS) grant to Y.T. and Y.C. and a Grant-in-Aid for Creative Scientific Research (19GS0211) to N.O.

275

435

440

445

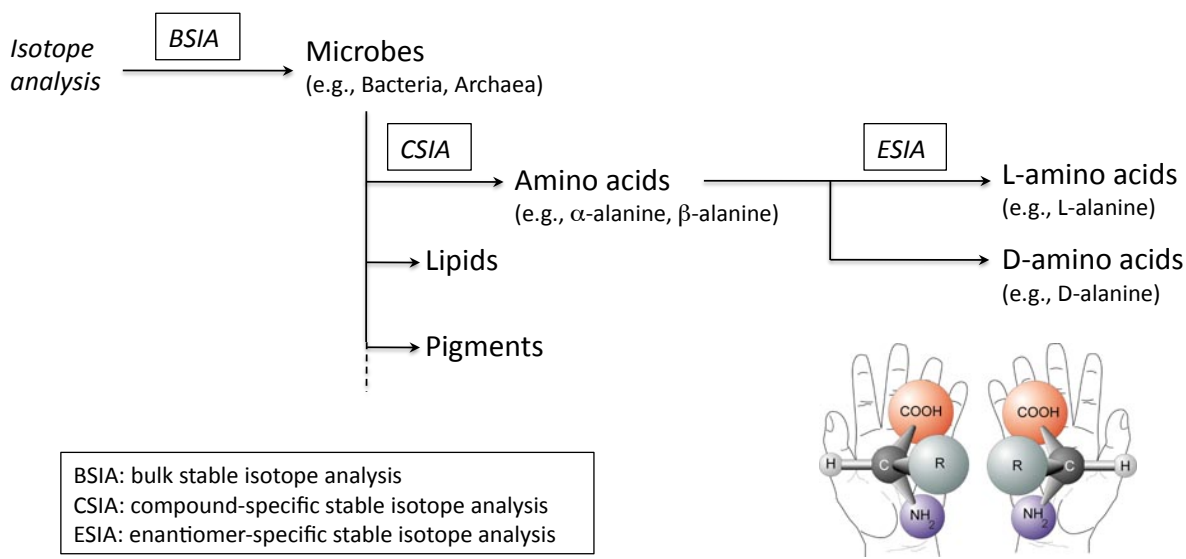

**Figure 1**

Elucidation schemes from bulk stable isotope analysis (BSIA), compound-specific stable isotope analysis (CSIA), and enantiomer-specific stable isotope analysis (ESIA). The word "Enantiomer-specific" studies are now well recognized as novel approaches to verify its origin, flux, and distribution (see Janak et al., 2005; Wedyan et al., 2008). The hand illustrate is courtesy by NASA (http://nai.arc.nasa.gov/library).

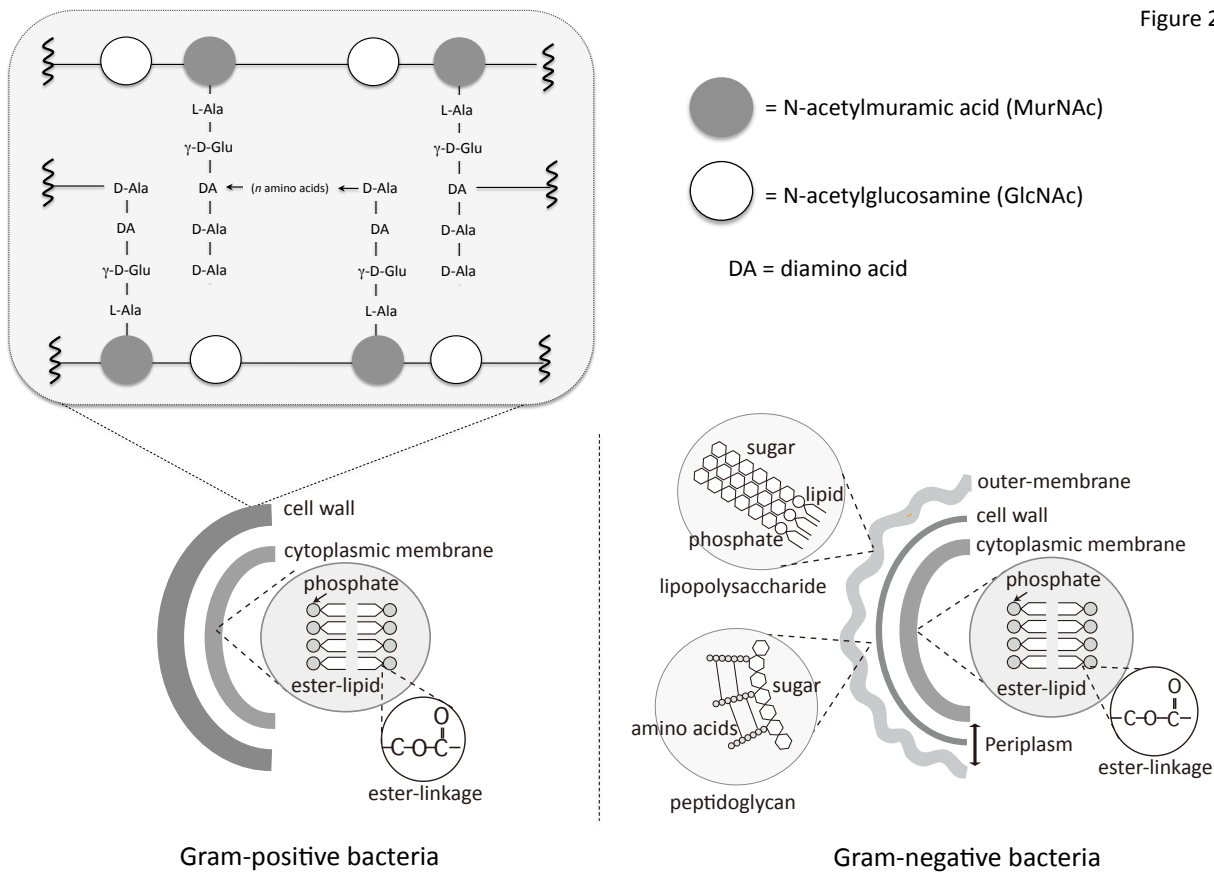

**Figure 2**

Primary structure of bacterial peptidoglycan and cell-wall in Gram-positive and Gram-negative Bacteria.

Abbreviations: GlcNAc: *N*-acetylglucosamine; MurNAc: *N*-acetylmuramic acid; DA: diamino acid (generally diaminopimelic acid or L-lysine); *n*: number of amino acids in the cross-bridge depending on the organism; →: CO–NH–. Modified from van Heijenoort, 2001a, 2001b.

**(a)**

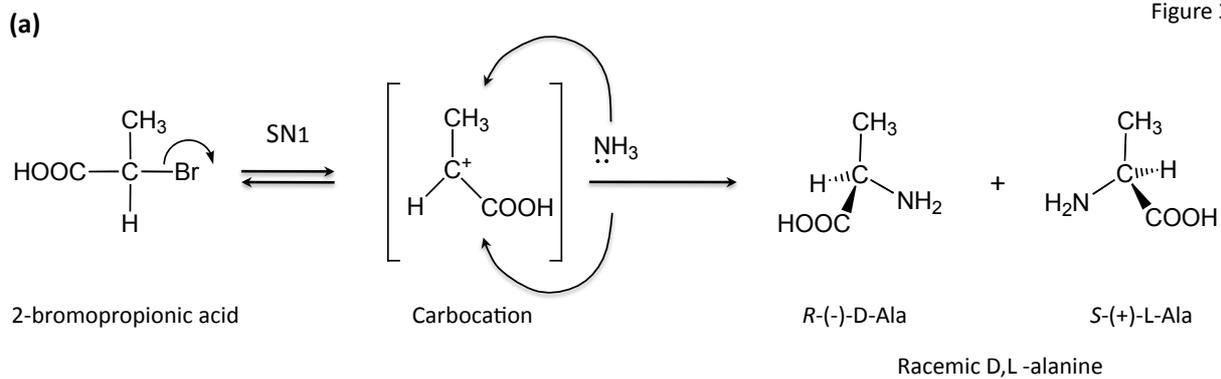

**(b)**

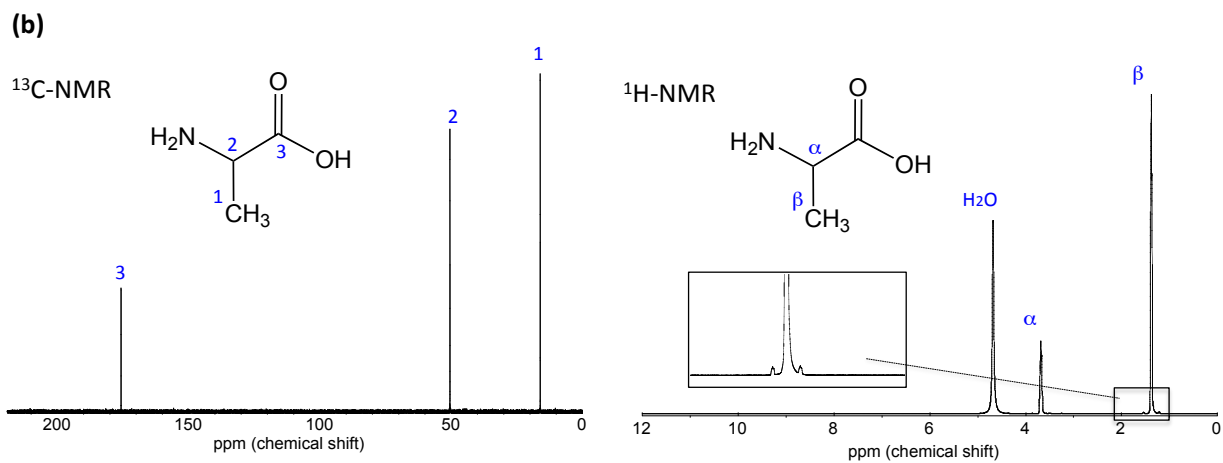

**Figure 3**

Schematic reaction of abiotic synthesis for racemic D- and L-alanine from 2-bromopropionic acid and ammonia by nucleophilic substitute 1 reaction pathway (SN1) *via* intermediate carbocation.

Full scan spectrum of $^{13}C$-NMR and $^{1}H$-NMR of synthesized D- and L-alanine. The satellite signals derived from natural abundance of $^{13}C$ on methyl-carbon ($CH_3$-) were appeared besides the main signal in $^{1}H$-NMR.

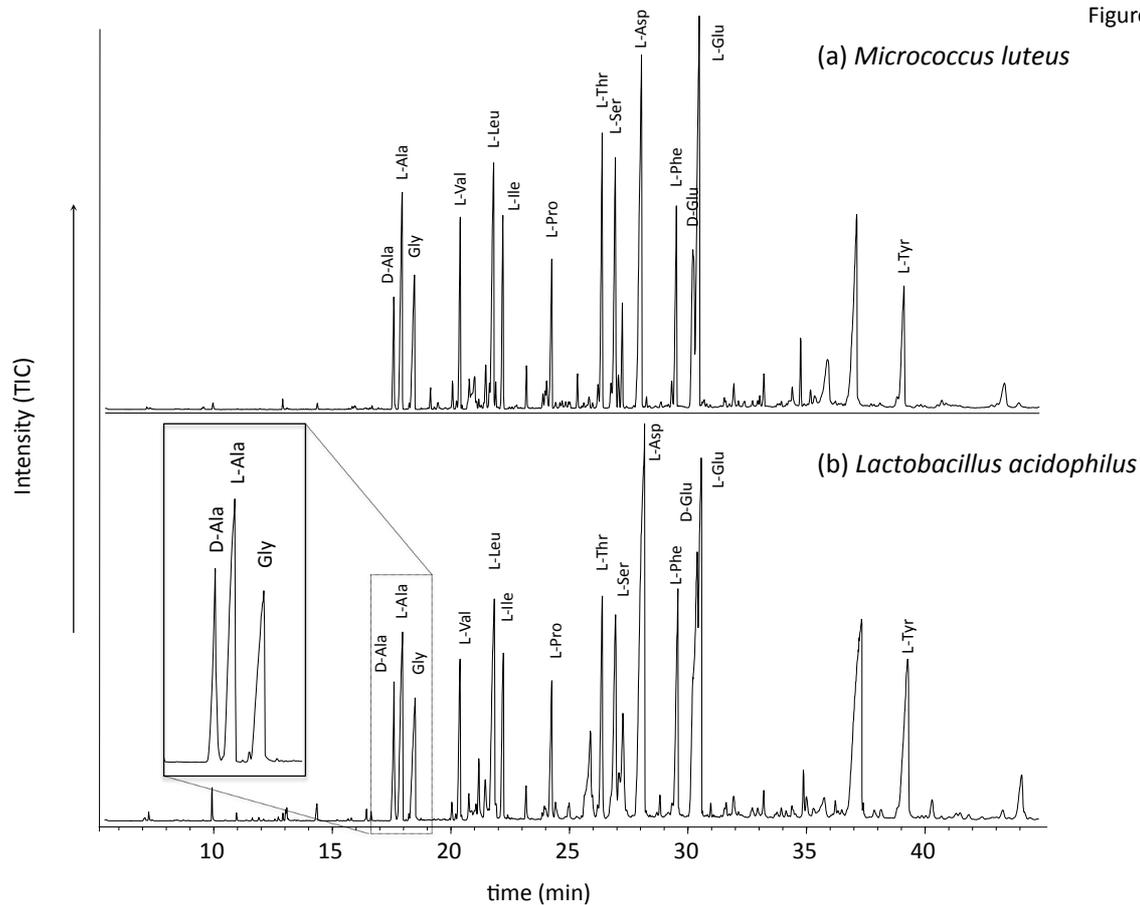

**Figure 4**

Representative chromatograms of GC/MS for D- and L-amino acids in peptidoglycan.

(a) *Micrococcus luteus* and (b) *Lactobacillus acidophilus*. Since GC/C/IRMS system has GC separation, on-line oxidation, on-line reduction, on-line $CO_2$ trap, and isotope mass detections, elution orders are same each other but the differences of retention time are inevitable by the differences of whole flow pathway (Takano et al., 2009).

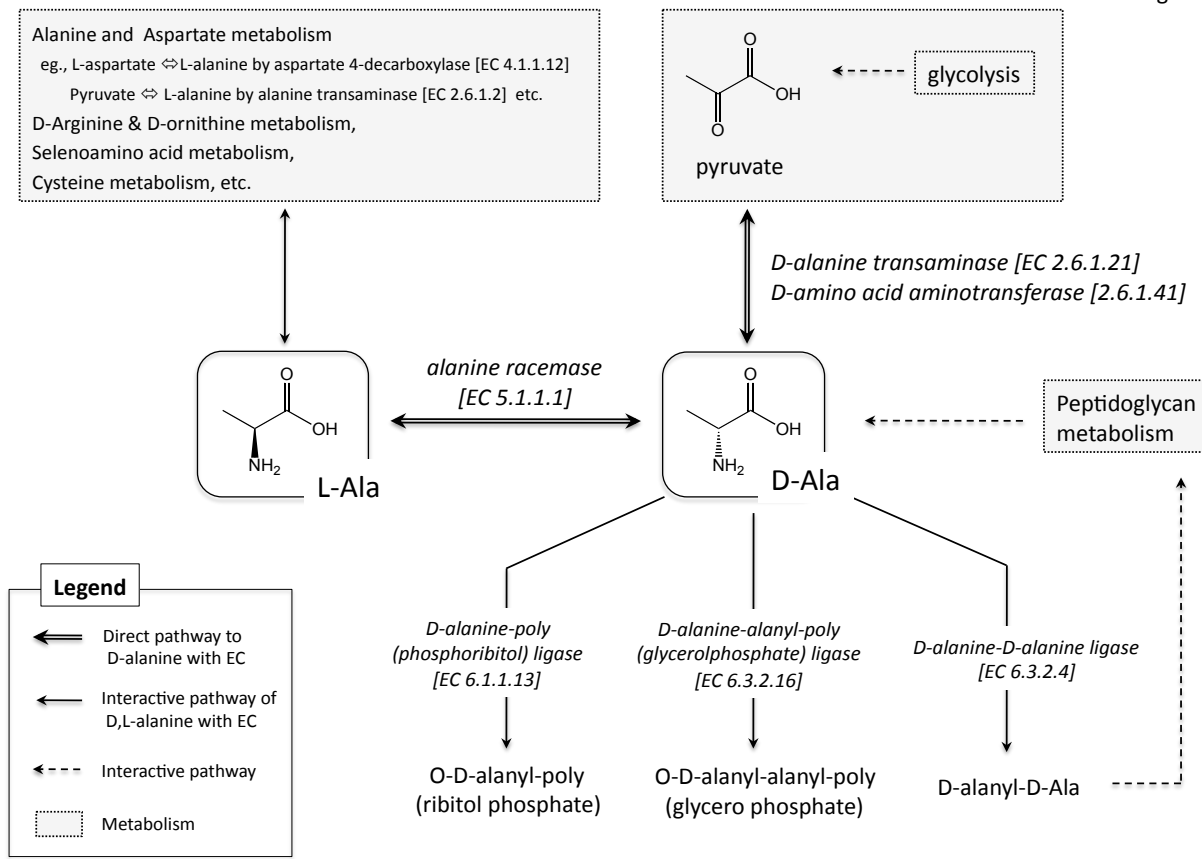

**Figure 5**

Pathway of D-alanine for incorporation to peptidogycan metabolism. We showed the main input pathway for L-alanine from Alanine and Aspartate metabolism, D-Arginine & D-ornithine metabolism, Selenoamino acid metabolism, and Cysteine metabolism. However some organisms has other input pathway for L-alanine from Taurine and hypotaurine metabolism, Carbon fixation, Reductive carboxylate cycle, and Vitamine B6 metabolism (see Kanehisa et al., 2006, Okuda et al., 2008, Kanehisa et al., 2008).

Enzyme Commission (EC) stands for enzyme genes to metabolic pathways (see also Schomburg et al., 2004: Barthelmes et al., 2007).

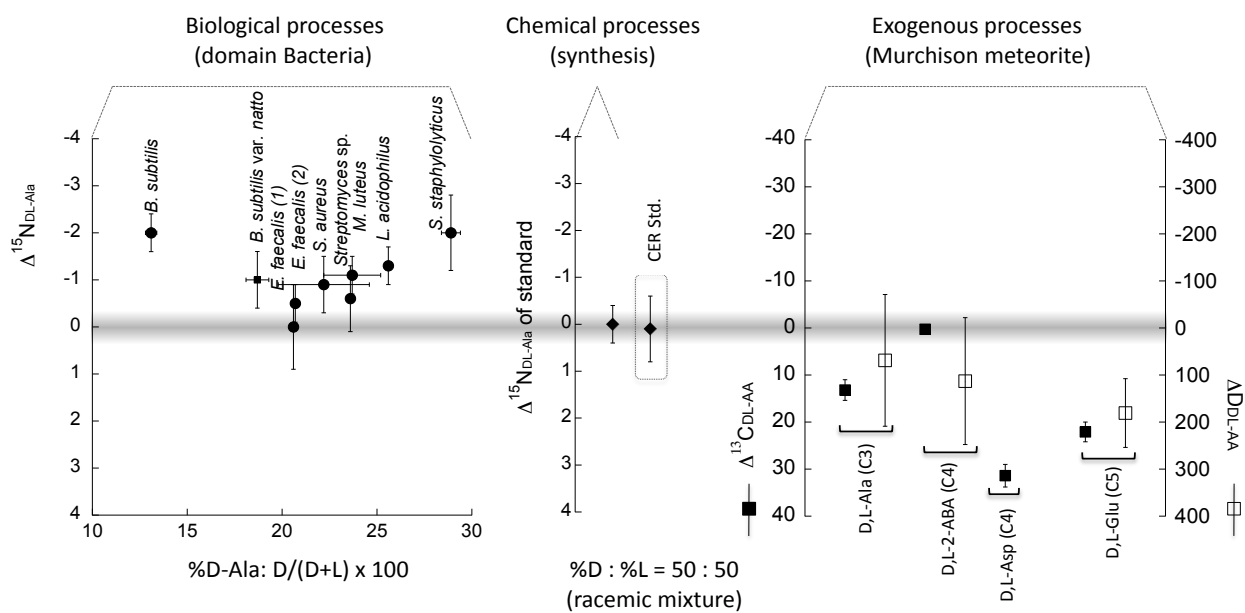

**Figure 6**

Comparison of $\Delta^{15}N_{D-L}$ ($\delta^{15}N_{D-Ala}$ - $\delta^{15}N_{L-Ala}$ ; defined as difference between $\delta^{15}N_{D-Ala}$ and $\delta^{15}N_{L-Ala}$) in biological process (e.g. domain Bacteria), chemical process (e.g. Organic symmetric chemical synthesis) with $\Delta^{13}C_{D-L}$ ($\delta^{13}C_{D-amino\ acid}$ - $\delta^{13}C_{L-amino\ acid}$ ; defined as difference between $\delta^{13}C_{D-amino\ acid}$ and $\delta^{13}C_{L-amino\ acid}$) and $\Delta D_{D-L}$ ($\delta D_{D-amino\ acid}$ - $\delta D_{L-amino\ acid}$ ; defined as difference between $\delta D_{D-amino\ acid}$ and $\delta D_{L-amino\ acid}$) of D- and L-alanine (C3; Carbon number 3), D- and L-aminobutyric acid (C4), D- and L-aspartic acid (C4), and D- and L-glutamic acid (C5) in exogenous process (e.g. Murchison meteorite, ESIA data after Pizzarello et al., 2004: Pizzarello and Huang, 2005).

The error bar for $\Delta^{15}N$ is based on the maximum standard deviations in $\delta^{15}N_{D-Ala}$ and $\delta^{15}N_{L-Ala}$ (‰ vs. Air scale), likewise in $\Delta^{13}C_{D-L}$ (‰ vs. PDB scale) and $\Delta D_{D-L}$ (‰ vs. SMOW scale). Black circles and squares in $\Delta^{15}N_{D-L}$ diagram represent peptidoglycan (Firmicutes and Actinobacteria: *Enterococcus faecalis*, *Staphylococcus aureus*, *Staphylococcus staphylolyticus*, *Lactobacillus acidophilus*, *Bacillus subtilis*, *Micrococcus luteus*, and *Streptomyces* sp.) and natural whole bacterial cells (*Bacillus subtilis* var. *natto*), respectively. Black diamonds in $\Delta^{15}N_{D-L}$ diagram stands represent chemical synthesize racemic mixtures of alanine (this study).

Abbreviated CER std., a working standard in Center of Ecological Research, Kyoto University was also compared in $\Delta^{15}N_{D-L}$ diagram. Black squares and open squares in exogenous process represent $\Delta^{13}C_{D-L}$ and $\Delta D_{D-L}$, respectively.

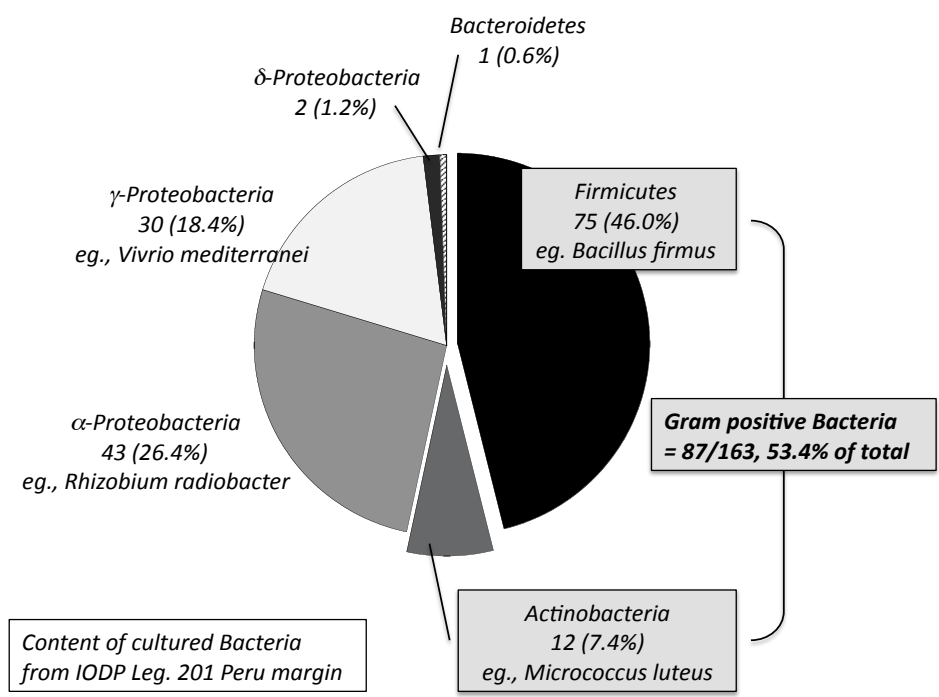

**Figure 7**

Compilation of cultured bacterial isolates from Ocean Drilling Program Leg 201 sediments in Open Pacific sites and Peru margin sites. See 16S rRNA sequence similarity and table data in D'Hondt et al., 2004, species listed are type species from GenBank database. In contrast, 16S rDNA clone libraries from ODP Leg 201 sites were dominated by uncultured lineages of Bacteria and Archaea (D'Hondt et al., 2004: Biddle et al., 2008). Although global intact polar lipids (IPLs) studies suggested significant dominant of Archaea in sub-seafloor (e.g. Biddle et al., 2006: Lipp et al., 2008), almost sub-surface Archaea has been still uncultured at present (e.g. Teske, 2008).

**Table 1**

Bacteria and their enzyme commission concerning with the D-alanine metabolism. Remarks stand for name of each culture inventory, and *Staphylococcus staphylolyticus* and *Micrococcus luteus* have been undescribed (see Kanehisa et al., 2006, Okuda et al., 2008, Kanehisa et al., 2008).

Abbreviations of class for direct input pathway to D-alanine; EC 5. Isomerases; 5.1 Racemases and epimerases; 5.1.1 Acting on amino acids and derivatives; 5.1.1.1 alanine racemase. EC 2. Transferases; 2.6 Transferring nitrogenous groups; 2.6.1 Transaminases; 2.6.1.21 D-amino-acid transaminase; 2.6.1.41 D-amino acid aminotransferase. Class for interactive pathway of D-alanine; EC 6. Ligases; 6.3 Forming carbon-nitrogen bonds; 6.3.2 Acid-D-amino-acid ligases (peptide synthases); 6.3.2.4 D-alanine-D-alanine ligase. EC 6. Ligases; 6.1 Forming carbon-oxygen bonds; 6.1.1 Ligases forming aminoacyl-tRNA and related compounds; 6.1.1.13 D-alanine-poly(phosphoribitol) ligase. EC 6. Ligases; 6.3 Forming carbon-nitrogen bonds; 6.3.2 Acid-D-amino-acid ligases (peptide synthases); 6.3.2.16 D-alanine-alanyl-poly(glycerolphosphate) ligase.

| Enzyme | Alanine racemase | D-alanine transaminase | D-amino-acid aminotransferase | D-Alanine-D-alanine ligase | D-Alanine-poly (phosphoribitol) ligase | D-Alanine-alanyl-poly(glycerolphosphate) ligase | Remarks |
|---|---|---|---|---|---|---|---|
|  | EC 5.1.1.1. | EC 2.6.1.21 | EC 2.6.1.41 | EC 6.3.2.4 | EC 6.1.1.13 | EC 6.3.2.16 |  |
| **Bacteria** |  |  |  |  |  |  |  |
| *Enterococcus faecalis* | + | - | - | + | + | - |  |
| *Staphylococcus aureus* | + | + | - | + | + | - | N315 |
| *Staphylococcus aureus* | + | + | - | + | + | - | JH1 |
| *Staphylococcus aureus* | + | - | - | + | + | - | JH9 |
| *Lactobacillus acidophilus* | + | - | - | + | + | - |  |
| *Bacillus subtilis* | + | + | - | + | + | - |  |

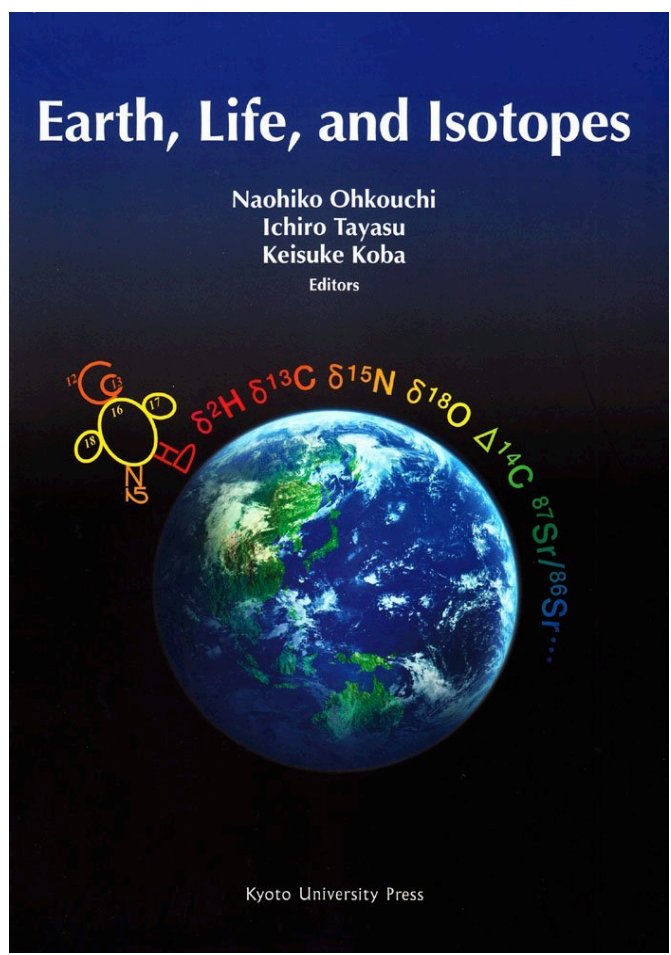